\documentclass[a4paper,fleqn]{cas-dc}

\usepackage[authoryear]{natbib}
\usepackage{amsmath}

\begin{document}
\let\WriteBookmarks\relax
\def\floatpagepagefraction{1}
\def\textpagefraction{.001}
\shorttitle{A novel early dark energy model}
\shortauthors{L.A. Garc\'ia et~al.}

\title{A novel early Dark Energy model} 

\author[1,2]{Luz \'Angela Garc\'ia}[orcid=0000-0003-1235-794X]
\address[1]{lgarciap@ecci.edu.co}
\address[2]{Universidad ECCI, Cra. 19 No. 49-20, Bogot\'a, Colombia, C\'odigo Postal 11131}
\author[3]{Leonardo Casta\~neda}
\author[3]{Juan Manuel Tejeiro}
\address[3]{Observatorio Astron\'omico Nacional, Universidad Nacional de Colombia}

\begin{abstract}
We present a theoretical study of an early dark energy (EDE) model. The equation of state $\omega(z)$ evolves during the thermal history in a framework of a Friedmann-Lemaitre-Robertson-Walker Universe, following an effective parametrization that is a function of redshift $z$. We explore the evolution of the system from the radiation domination era to the late times, allowing the EDE model to have a non-negligible contribution at high redshift (as opposed to the cosmological constant that only plays a role once the structure is formed) with a very little input to the Big Bang Nucleosynthesis, and to do so, the equation of state mimics the radiation behaviour, but being subdominant in terms of its energy density. At late times, the equation of state of the dark energy model asymptotically tends to the fiducial value of the De Sitter domination epoch, providing an explanation for the accelerated expansion of the Universe at late times, emulating the effect of the cosmological constant. The proposed model has three free parameters, that we constrain using SNIa luminosity distances, along with the CMB shift parameter and the deceleration parameter calculated at the time of dark energy - matter equality. With full knowledge of the best fit for our model, we calculate different observables and compare these predictions with the standard$\Lambda$CDM model. Besides the general consent of the community with the cosmological constant, there is no fundamental reason to choose that particular candidate as dark energy. Here, we open the opportunity to consider a more dynamical model, that also accounts for the late accelerated expansion of the Universe.
\end{abstract}


\maketitle

\section{Introduction}
Observations of the luminosity distances of the Supernova type Ia \citep[SNIa;][]{riess2000} revealed that the expansion of the Universe is speeding up at late times. Within the cosmological standard model, there is an unknown matter--energy component that contributes by about 70\% of the critical density, and this fluid is described as a smooth component with negative pressure. Although astronomers know the effect of this fluid, there is not a clear idea of how to detect it, mainly because it is a smooth component, dilute throughout all the Universe and the parameter of the equation of state today is most likely $\omega_0 = -1$, even if $\omega = \omega(t)$ in the past.\newline

Different models have been proposed in the past years to explain the nature of this component: the cosmological constant $\Lambda$ that accounts for the quantum va\-cuum energy \citep{carroll2001,peebles2003}, scalar fields with different $\omega(t)$: Quintessence fields \citep{ratra1988,caldwell1998,sami2003} (with the state equation $\omega = \frac{p_{Q}}{\rho_{Q}} =$ constant), K--essence \citep{armendariz1999,chiba2000,armendariz2001,chiba2002}, Taquionic fields \cite{sen2002a,sen2002b,gibbons2002}, phantom fields \citep{caldwell2002,cline2004}, frustrated topological defects, extra--dimensions, massive (or massless) fermionic fields, galileons, effective pa\-ra\-me\-tri\-za\-tions of the state equation, primordial magnetic fields, Chaplygin gas \citep{kamenshchik2000,bento2004}, holographic models \citep{horava2000}, Horndeski's theory \citep{clifton2012}, and, early dark energy models \citep{wetterich2004,doran2006,ede2020}, among others. All these models can be predicted by the Friedmann equations in the framework of General Relativity. Instead, modified gravity models impose the accelerated expansion through a geometrical contribution, rather than an energy density \citep{capozziello2010,felice2010}.\newline

The current paradigm in the standard model is the $\Lambda$CDM model, that has only a few free parameters, well-constrained with present observations. Nonetheless, the nature of the cosmological constant $\Lambda$ is still unexplained. One can wonder if it is not more natural that the accelerated expansion could have been produced by a different smooth field, that evolves with redshift $z$, having a non-null contribution in the early Universe and emulating the action of the cosmological constant $\Lambda$ at late times. \newline

During the radiation domination epoch, the abundances of light nuclei predicted by the Big Bang Nucleosynthesis \citep[BBN][]{alpher1948,gamow1946}, in particular $Y_{He}$, can be used to quantify the degrees of freedom of the radiation components in the early Universe \citep{garcia2011}. At the matter domination era, when matter components are predominant and structure was formed, baryonic acoustic oscillations (BAO) and the anisotropies in temperature of the Cosmic Microwave Background also allow astronomers to constrain their dark energy (DE) models, although DE is not the main contributor of the matter-energy density in that stage.\newline 

Additional tests, such as the calculation of the age of the Universe or the \textit{Statefinder} parameters can tell us the deviation from a given DE model from the $\Lambda$CDM predictions, and how feasible a DE candidate is in the observed Universe. \newline
With the aim to give a plausible explanation of the accelerated expansion of the universe, we have proposed a model of dark energy which has a non--null contribution at early times to increase the Hubble radius during radiation domination era and influence the Boltzmann equations that determine the evolution of the light abundances. All the conditions that allow us to describe the early dark energy are achieved with the effective parametrization, which is characterized by its equation of state that mimics the dominating component. \newline

Throughout the paper, we use the cosmological parameters from the Planck Collaboration \citep{planckcol2018} with $\Omega_{0m}=$ 0.3111 $\pm$ 0.0056, $\Omega_{\Lambda}=$ 0.6889 $\pm$ 0.0056 and $H_0=$ 67.66 $\pm$ 042 km s$^{-1}$Mpc$^{-1}$ (or $h =$ 0.6766), and a spatially--flat model of the Universe with cold dark matter. \newline

The paper is presented as follows: in Section~\ref{sec:2}, we des\-cri\-be an alternative DE candidate with a non-negligible contribution in the early Universe and that mimics the cosmological constant $\Lambda$ effect at late times. Section~\ref{sec:3} shows the method employed to find the best fitting parameters of the model proposed as a different option to dark en al. In Section~\ref{sec:4}, we explore the evolution with redshift of the dark energy density fraction and compare the observables predicted by our model with current cosmological observations. Section~\ref{sec:5} discusses proxies that the model is submitted to constrain it and compare it with the current paradigm, $\Lambda$, in the standard model. Finally, Section~\ref{sec:6} summarizes the findings of this study and proposes perspectives for DE candidates, which evolution with redshift is not well constrained with observations yet. 

\section{Effective parametrization of $\omega_{\phi}$}
\label{sec:2}
The goal of this work is to study a theoretical prescription that describes dark energy. In order to do so, we make no assumption on the nature of the dark energy component in our model, hence, it can be described as a non-interacting perfect fluid that evolves with other components of the plasma in the Universe, and therefore, it could be supposed as scalar field $\phi$. We do not discuss further the nature of the DE model but leave open the possibility to link it with a particular scalar field in the literature. \newline
The prospective model takes into account that the Universe is experiencing an accelerated expansion at late times, following a De Sitter attractor, hence $-1< \omega_{de} < - \frac{1}{3}$. As a consequence, it should be compared with the cosmological constant $\Lambda$. Ultimately, the motivation of this work is to find the physical insight of the current expansion of the Universe, but also to give an alternative to $\Lambda$CDM model, since there is a significantly large difference with the energy of vacuum fluctuations.\newline
In addition, we are interested to establish a realization that has a non-negligible contribution during the radiation domination epoch, being subdominant with respect to the radiation energy fraction. Therefore, our model could have an input to BBN, through effective degrees of freedom introduced in the Hubble parameter $H(z)$, as long as this action does not overcome the observational upper limits. \newline
A contribution at high redshift ($z \sim$ 10$^{11}$) can be achieved by imposing $\omega_{\phi}=1/3 \vert_{v_{rad}}$, and the following condition for our  DE' energy density:
\begin{equation}\label{hungry}
\rho_{de}\vert_{rad}=b \cdot \rho_{rad} \:\:\:\:\:\:\:\:\:\:\:\:\:\:\: 0\leq b < 1.
\end{equation}
\noindent with $\rho_{rad}$ the radiation energy density. The equation~\eqref{hungry} can be described through the assumption  $\rho_{de}\vert_{rad} \propto a^{-4}$. As a result, the early dark energy model would be characterized by an effective parametrization, that evolves in time, without impacting the hierarchy and chronology of the events in the cosmic history. The parametrization of the equation of state $\omega_{de}$ should converge to the limits mentioned above.\newline

Different parametrizations have been proposed to describe the evolution of DE and/or a unified dark matter and dark energy model \cite{davari2018} or constraints at high redshift \cite{lorenz2017}. In order to achieve a general solution of the dynamical system established in previous section, we propose an effective parametrization of the state equation valid up to very high redshift, towards to the Planck time:
\begin{equation}\label{fun}
\omega_{\phi}(z) = \frac{4/3}{\left(\frac{1+z_{*}}{1+z}\right)^m+1} - 1.
\end{equation}
\noindent Here, $m$ is a factor that modules the transitions between the attractors, $z_{*}$ is a redshift in matter domination epoch defined by:
\begin{equation}
z_{*} = \frac{z_{eq}+z_{de}}{2}
\end{equation}
\noindent with $z_{eq}$, the matter--radiation equality and the $z_{de}$, the redshift when the De-Sitter domination (i.e. the accelerated expansion of the Universe stage) begins.\newline

The parametrization \eqref{fun} respects all the conditions previously mentioned, thus, explores an alternative to the $\Lambda$CDM current paradigm. \newline

The energy density of the dark energy component $\rho_{\phi}$ is given by:
\begin{equation}\label{intrho}
\int_{\rho}^{\rho_0} \frac{d\rho^{\prime}}{\rho^{\prime}} = - 3\int_{a}^{1} \frac{(1+\omega_{\phi}(a^{\prime}))}{a^{\prime}}da^{\prime},
\end{equation}
\noindent integrating \eqref{intrho}, it is obtained:
\begin{equation}\label{rhoend}
\rho =\rho_0 \cdot (1+z)^{4} \left[\frac{\left(\frac{1+z_{*}}{1+z}\right)^m+1} {\left(1+z_{*}\right)^m+1} \right]^{4/m}=\rho_0 \cdot f(z).
\end{equation}
\noindent with:
\begin{equation}\label{fa}
f(z) = (1+z)^{4} \left[\frac{\left(\frac{1+z_{*}}{1+z}\right)^m+1} {\left(1+z_{*}\right)^m+1} \right]^{4/m}.
\end{equation}
\noindent Moreover, the fraction of the dark energy density $\Omega_{\phi}=1-F=\frac{\rho_{\phi}}{\rho_{cr}}$:
\begin{equation}\label{frack}
\Omega_{\phi}(z)=\frac{\rho_0}{\rho_{\text{cr}}}=\frac{\Omega_{\phi 0} \cdot f(z)}{\Omega_{\phi 0}\cdot  f(z) + \Omega_{m 0} \cdot  (1+z)^{3}}.
\end{equation}
\noindent We remind the reader that we assume a spatially--flat Universe and a Concordance model, hence, $\Omega_{\phi 0} + \Omega_{m 0} + \Omega_{\text{rad} 0} = 1$ and $\Omega_{\text{rad} 0} \rightarrow$ 0 at late times.

\section{Best fitting parameters of the model}
\label{sec:3}

The formal solution of the parametrization \eqref{fun} requires the estimation of the free parameters of the model $\{\Omega_{\phi_0}, m, z_{de}\}$, the fraction of the dark energy density, the module that regulates the transition between the radiation to the De-Sitter domination eras, and the dark energy domination redshift, respectively. A preliminary inspection of the parameter-space shows a large degeneracy between $m$ and $z_{de}$.\newline

We use observations of the luminosity distances of SNIa from the survey \textsc{Supernova Cosmology Project Union2.1} \citep[SCP2.1][]{rubin2014}\footnote{\url{http://www-supernova.lbl.gov/}}, along with the CMB shift parameter $R_{\text{CMB}}$ and, the condition of the deceleration parameter equals to zero at $z = z_{de}$. Adopting $R_{\text{CMB}}$ in this work, allow us to constrain the free parameters of the model at high redshift, during the matter domination era.\newline

We build an MCMC module to find the set of best-fitting parameters to the model. The priors of the model proposed can be summarized as:
\begin{itemize}
\item $\Omega_{\phi_0}$ should be strictly positive, $[0,1]$ in the Concordance model.
\item  Negative values of $m$ lead to an inverted transition between the radiation and the De-Sitter attractors (the latter occurring first than the former), which is not consequent with the thermal history of the Universe. On the other hand, $m =$ 0 produces no transition whatsoever, then, $m$ is strictly positive in the framework of the Standard Model. Furthermore, visual inspection of the evolution of this parameter shows that $m >$ 90 leads to a quick transition (for very large values of $m$ to an instantaneous transition) between the attractors. We discard these values of $m$ because they are unlikely from the observational point of view. In fact, the structure was formed during the matter domination epoch, which would not happen if there would not have existed an extended transition between the radiation and De-Sitter domination eras. 
\item The redshift of matter -- dark energy equality, $z_{de}$ has already occurred since the Universe is experiencing an accelerated expansion $\Rightarrow$ 0 $ < z_{de} \gtrsim$ 1.5. The upper limit takes into account that cosmic structure was formed during the matter domination epoch, and that has been observed through different with different surveys to-date \textsc{2dFGRS} \footnote{\url{http://www.2dfgrs.net/}}, \textsc{6dFGS} \footnote{\url{http://www.6dfgs.net/}}, \textsc{WiggleZ} \footnote{\url{http://wigglez.swin.edu.au/}} and the Sloan Digital Sky Survey \textsc{SDSS} \footnote{\url{https://www.sdss.org/}}. 
\end{itemize}

We must break the large degeneracy between $m$ and $ z_{de}$. To do such, as well as to find the best fitting values for $\{\Omega_{\phi_0}, m, z_{de}\}$ set, we use the luminosity distance $d_L (z)$ -equation~\eqref{dl} - and the distance modulus $\mu$ -equation~\eqref{mu}- are built for our model to compare these functions with the observational SNIa distance modulus from SCP2.1, with $z$ up to 1.4.
\begin{equation}\label{dl}
d_L (z)= \frac{c (1+z)}{H_0}\int_{0}^{z}\frac{dz^{\prime}}{B(z^{\prime})},
\end{equation}
\begin{equation*}
B(z^{\prime})=(\Omega_{\phi_0}f(z^{\prime};m, z_{*})+(1-\Omega_{\phi_0})(1+z^{\prime})^{3})^{1/2},
\end{equation*}
\begin{equation}\label{mu}
\mu=m-M= 5 (\text{log}_{10}d_L(z)-1).
\end{equation}

As mentioned above, the CMB shift parameter $R_{\text{CMB}}$ is also imposed as condition to constrain the set of free parameters of the model. The function $R_{\text{CMB}}$ measures the shifting of the acoustic peaks from from the BAO \citep{bond1997,efstathiou1998} and it is defined as the comoving distance between the last scattering surface and today:\newline
\begin{equation}
R=(\Omega_m H_0^2)^{1/2}\int_0^{1089}\frac{dz}{H(z)},
\end{equation}
\noindent Neither \citet{planck2015} or \citet{planckcol2018} calculate directly the value of this parameter, different than the WMAP-7 that inferred the value of the CMB shift parameter as $R=1.719 \pm 0.019$ \citep{panotopoulos2011}. Nonetheless, \citet{huang2015} use cosmological parameters from \citet{planck2015} to compute an updated value of $R =$ 1.7496 $\pm$ 0.005. It is worth noticing that this is the only observable that we constrain with \citet{planck2015}, and not with\citet{planckcol2018} cosmological parameters, as stated in the introduction.\newline

\begin{figure*}
\centering
   \includegraphics[scale=0.4]{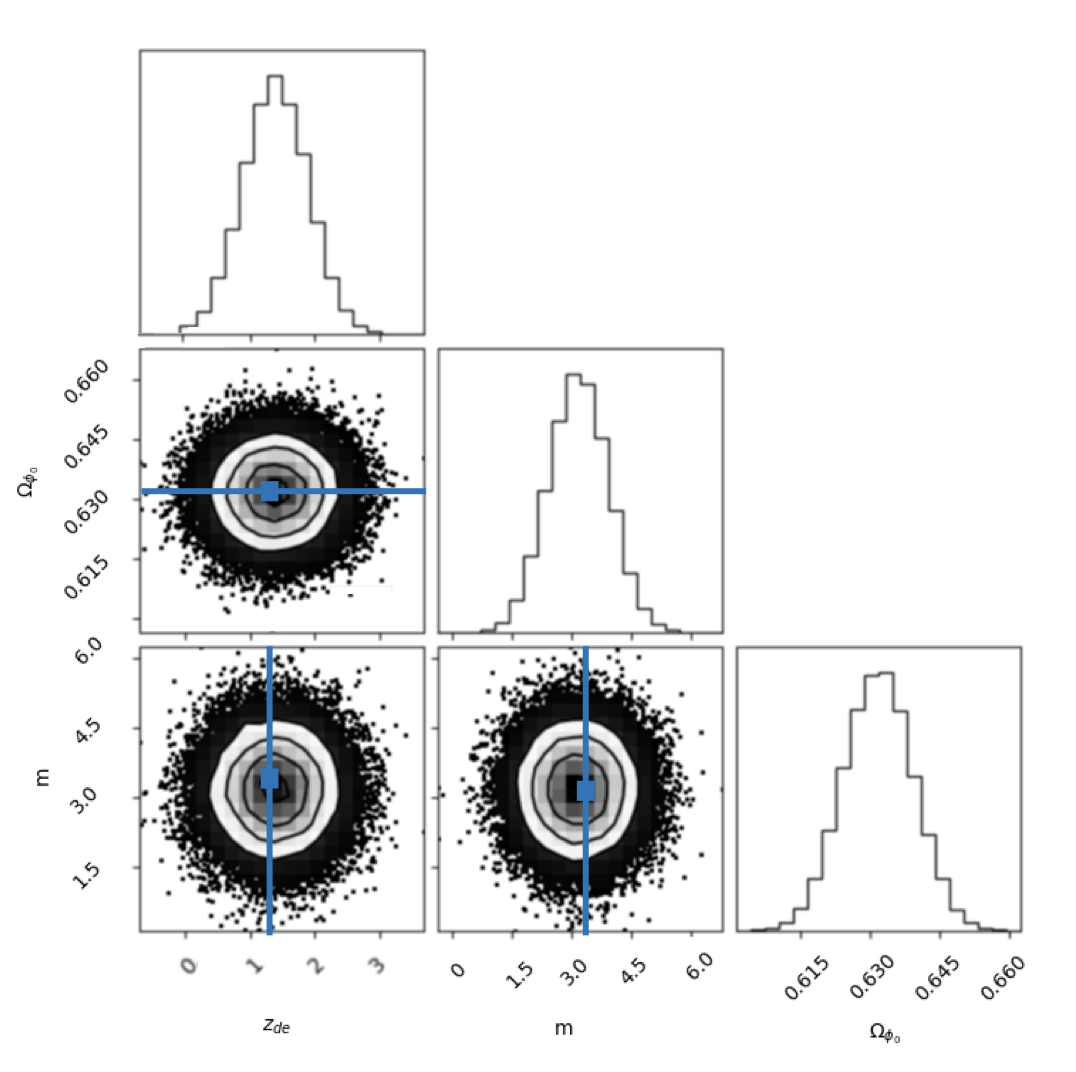} 
\caption{Posteriors of our free parameters $\Omega_{\phi_0}$, $m$ and $z_{de}$, with shaded 68\% intervals, fitting to SNIa luminosity distances data from SCP2.1, on top of the simultaneous constraints given by the CMB shift parameter $R_{\text{CMB}}$ and the deceleration parameter $q(z_{de})=0$. The best values estimated with the MCMC method lay within the prior conditions. With our analysis, we are able to recover the best values of the free parameters: $\Omega_{\phi_0} = $0.63 $\pm$ 0.05, $m =$ 3.2 $\pm$ 0.9, and $z_{de} =$ 1.2 $\pm$ 0.3. The latter parameters maximize the likelihood function, and break the tight degeneracy existent between $m$ and $ z_{de}$.}
\label{fig:lala}
\end{figure*}

The third condition assumed to calculate the three free parameters of the model is the deceleration parameter condition  $q(z_{de})=0$, i.e. the Universe starts it accelerated expansion at the time that the dark energy density overcomes other matter-energy contributions. Although the deceleration parameter is no longer used in the framework of the Concordance model, its definition and solution are quite handy to narrow down the degeneracy between $m$ and $ z_{de}$.\newline
\begin{equation}
q(z_{de})=0 \:\:\:\:\:\:\:\: \text{with} \:\:\:\:\:\:\:\: q(z)= (1+z)\frac{\partial H}{\partial z}-1.
\end{equation}
%
The best-fitting parameters are obtained with an MCMC module that takes into account the three conditions previously described. Figure~\ref{fig:lala} shows the posteriors of $\Omega_{\phi_0}$, $m$, and $ z_{de}$, and it has been calculated with 3 walkers in the MCMC routine built in python. It converges after 100000 steps around the parameter--space. The $R_{\text{CMB}}$ constrain at high redshift is determinant to break the degeneracy occurring in two of the three free parameters. The corner plot~\ref{fig:lala} shows the best fits to the early dark energy model in blue and then, the 68\% interval regions allow us to determine the errors of the model.\newline
The best estimates for the free parameters of the model and their errors are displayed in Table~\ref{tab:table}, as well as some derived parameters relevant to cosmology. We compare these best-fitting values with the ones from the $\Lambda$CDM model from \citet{planckcol2018}.\newline
\begin{table}[h!]
\centering
\caption{Summary of the best values of the free parameters of the DE model and comparison with the $\Lambda$CDM. Column 1: parameter name. Column 2: estimates for our model. Column 3: $\Lambda$CDM comparison \citep{planckcol2018}.}
\begin{tabular}{lcc} 
  \hline
\textbf{Parameter} &  \textbf{Our model} & \textbf{$\Lambda$CDM  model} \\ \hline 
$\Omega_{\phi_0}$ & 0.631 $\pm$ 0.005 & 0.6889 $\pm$ 0.0056 \\
$m$ & 3.2 $\pm$ 0.9 & -- \\ 
$z_{de}$ & 1.2 $\pm$ 0.3 & -- \\ 
 \hline
$\Omega_{m_0}$ & 0.369 $\pm$ 0.005 & 0.3111 $\pm$ 0.0056\\ 
$\omega_0$ & -0.976 $\pm$ 0.358 & -1\\ \hline
\end{tabular}
\label{tab:table}
\end{table}

Numerical calculations made with our background model, allow us to report the CMB shift parameter associated $R_{\text{cal}} =$ 1.85 $^{+\text{0.12}}_{-0.13}$. The value is slightly larger than the one calculated by \citet{huang2015}, but as claimed before, our model differs from $\Lambda$CDM result, as expected. Besides, $R$ depends strongly on the factor $H(z)$, that changes with the model considered. In addition, we remind the reader that we adopt the \citet{planck2015} cosmological parameters for this particular observable, rather than \citet{planckcol2018}, since the value has not been reported yet in the literature with the latter cosmological parameters.

\section{Evolution of the observables associated to the DE model}
\label{sec:4}

\begin{figure}
	\includegraphics[width=\columnwidth]{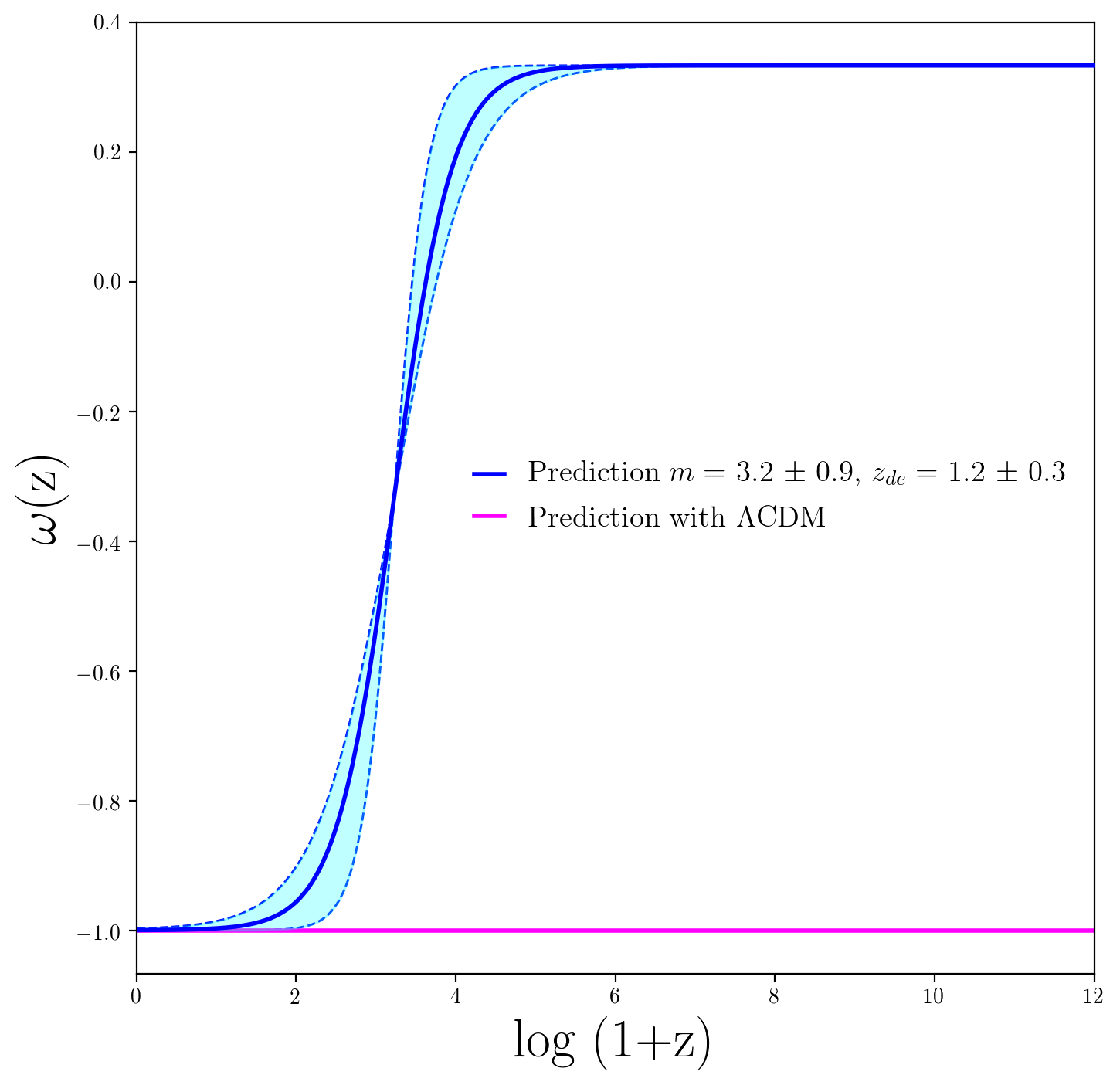}
\caption{Equation of state $\omega (z)$ as a function of redshift $z$. The parameterization is valid up to the Planck time. The blue line represents the equation of state of the field, the dashed lines show the upper and lower limits of $\omega (z)$, whereas the cyan region displays all the possible values that the equation of state could take inside the parameter-space allowed within the error bars. The purple line shows a comparison of the evolution of the equation of state in the case of the $\Lambda$CDM model.}
\label{fig:lolo}
\end{figure}

Once the set of free parameters has been constrained with the MCMC method, the evolution of the dark energy model is complete and can be studied in the different cosmological eras.\newline
Figure~\ref{fig:lolo} shows the evolution of the equation of state $\omega$ as function of $z$. The early DE model emulates radiation du\-ring this epoch and then, it evolves to the De-Sitter era. At late times, our theoretical description emulates the cosmological constant. In fact, dark energy relaxes to the asymptotic $\Lambda$CDM model during De-Sitter epoch. The blue curve shows the equation of state for our model, and the dashed lines represent the upper and lower limits imposed by the errors of the set of parameters $\{\Omega_{\phi_0}, m, z_{de}\}$  and the shaded cyan region display the possible range where $\omega (z)$ can evolve inside the error bars.\newline

\begin{figure}
	\includegraphics[width=\columnwidth]{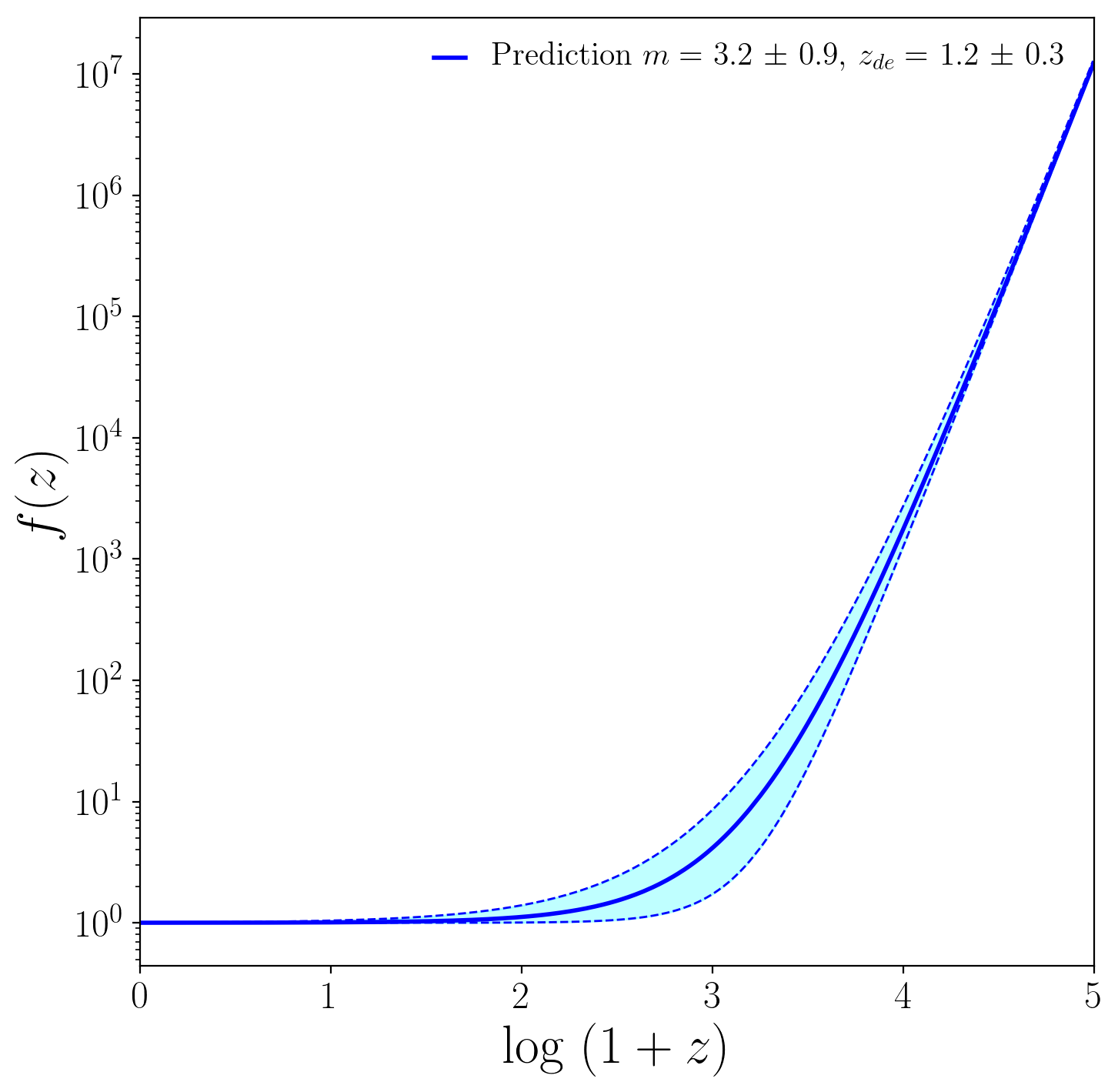}
\caption{Dark energy density factor $f(z)$ as a function of redshift $z$. The factor grows with redshift, differently from the evolution of the dark energy density $ \Omega_{\phi}$. The blue continuous line shows the evolution of $f(z)$, whereas the dashed lines the upper and lower limits of the function inside the error bars of the parameters. The cyan shaded region, all the possible values of $f(z)$ within the parameter-space.}
\label{fig:lulu}
\end{figure}

On the other hand, Figure~\ref{fig:lulu} presents the behaviour of $f(z)$. This function characterizes the evolution of the dark energy density in our model. During radiation domination epoch, the field scales as radiation $\rho \propto (1+z)^{4}$ until $z_{eq}$. After that, $\rho$ has a complex behaviour which guarantees the late convergence to accelerated expansion. At this point, the model evolves asymptotically to $-1$ (as the cosmological constant). At $z =$ 0, the function $f(z = 0) =$ 1, by construction, indicating that the dark energy density of the field is dominant over matter and the dark energy candidate satisfies the current observations and is in agreement with the predictions from the Concordance model.\newline

\begin{figure}
	\includegraphics[width=\columnwidth]{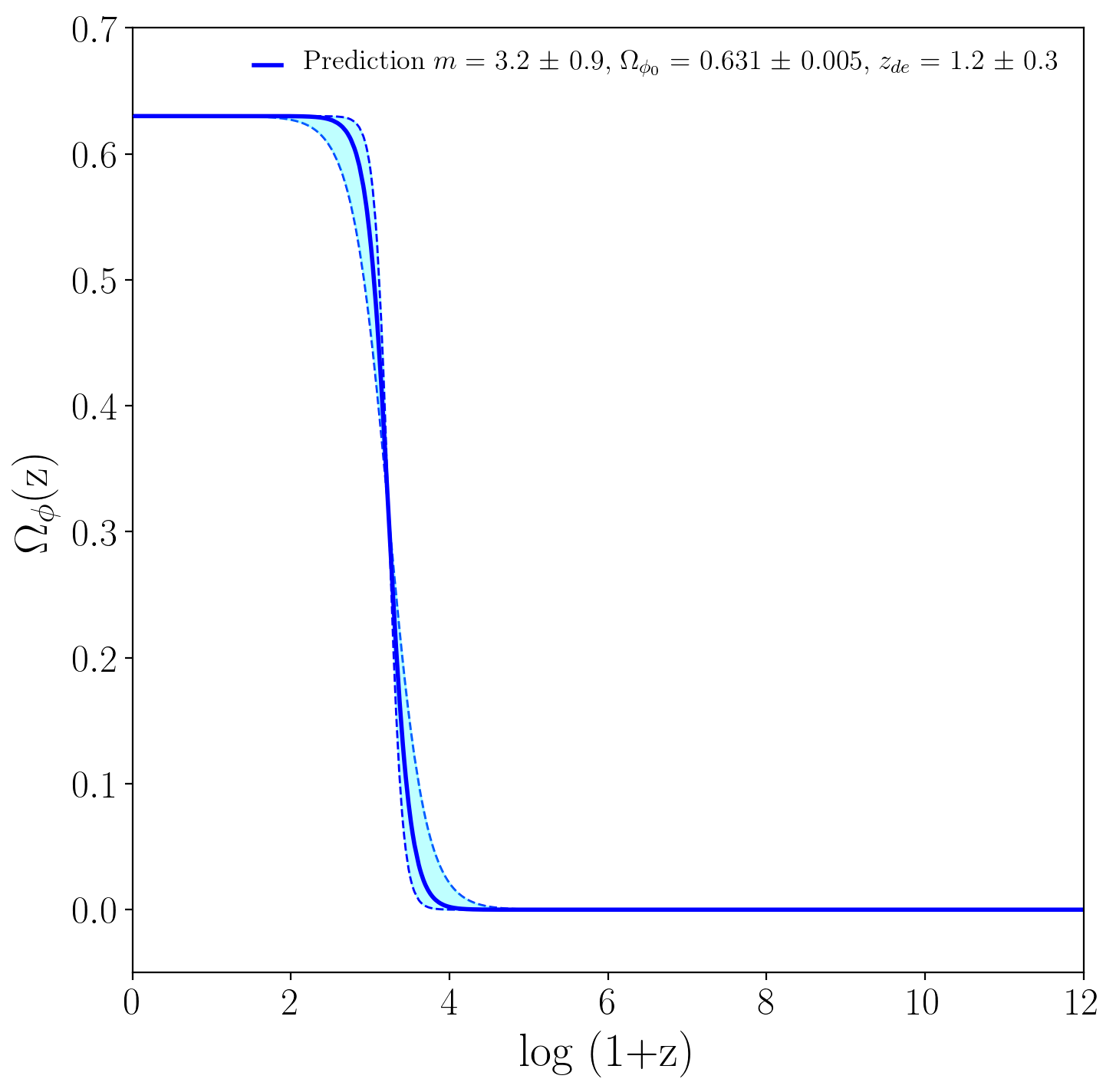}
\caption{Dark energy density fraction $\Omega_{\phi}$ with redshift $z$. The blue line presents the evolution of the energy density of the field. The dashed lines show the upper and lower limits of the dark energy density inside the error bars of the parameters. Within these boundaries, in the cyan shaded region, the possible values of $\Omega_{\phi}$ within the parameter-space. It is worth noticing that at $z \sim$ 10$^{10}$, $\Omega_{\phi}$ raises from a value close to zero, indicating that our model is an EDE and it energy density fraction could contribute with some effective degrees of freedom in the Hubble parameter at radiation domination era.}
\label{fig:ala}
\end{figure}

The evolution of the dark energy density fraction is shown in Figure~\ref{fig:ala}. In the plot, it is possible to distinguish that $\Omega_{\phi 0} =$ 0.631, the value of the dark energy density fraction today. When times evolves back (i.e. increasing $z$), the energy density of the field decreases, being subdominant during matter and radiation epoch, as imposed by construction with the parametrization of the equation of state. Nonetheless, the value of the energy density fraction of the model has a non-negligible contribution of the field energy density.\newline

\begin{figure}
	\includegraphics[width=\columnwidth]{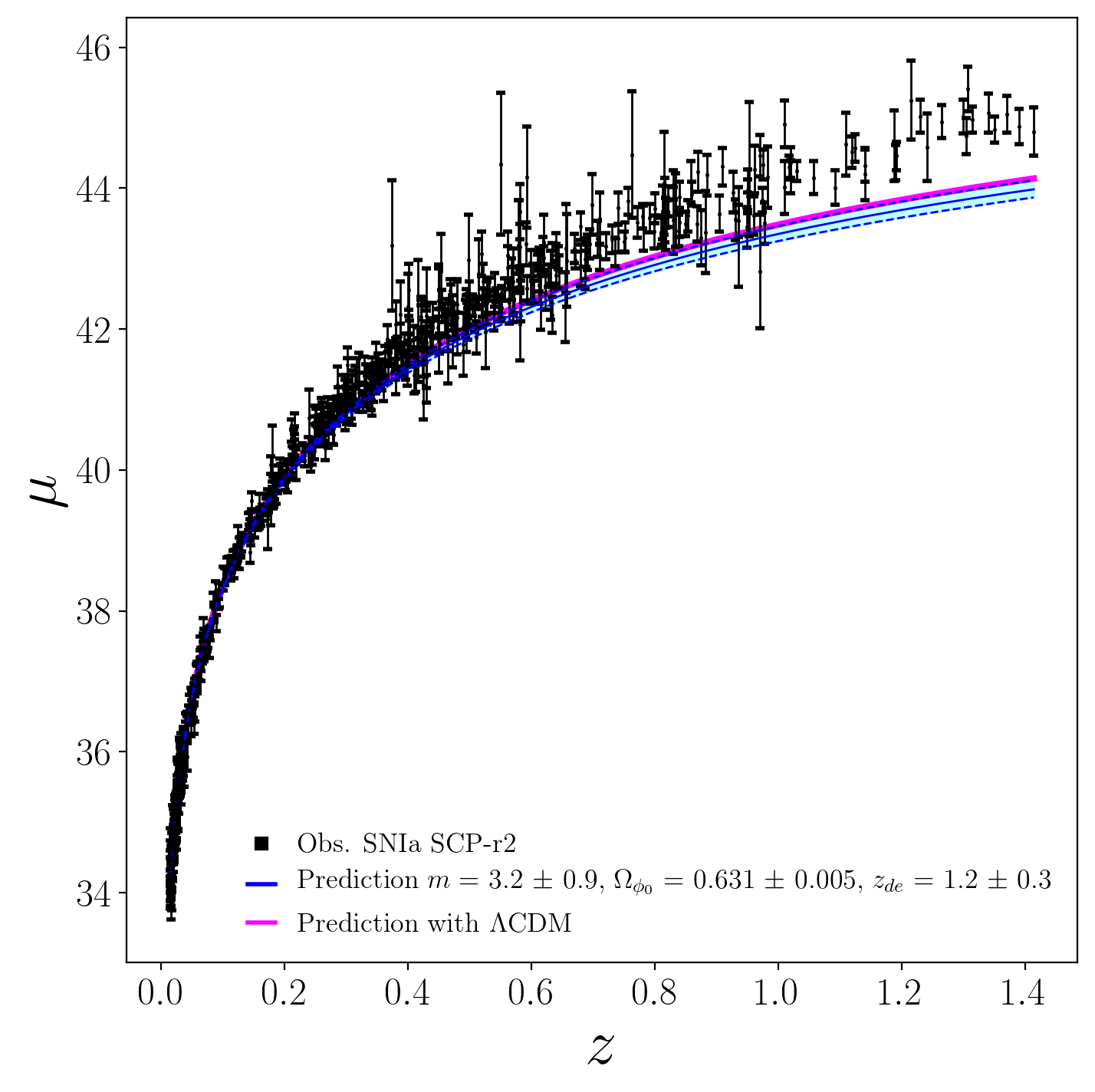}
\caption{Distance modulus vs. redshift $z$ computed with our model and $\Lambda$CDM. We compare the theoretical predictions with observational data of SNIa from SCP2.1. We present our model, $\Lambda$CDM and SNIa from SCP release in the blue line, magenta line and black points, respectively.}
\label{fig:ali}
\end{figure}

Moreover, we analyse the luminosity distance in our model with the best fitting parameters found in the previous section. Figure~\ref{fig:ali} displays the distance modulus in our model in a blue line (with the boundaries inside the parameter-space in blue dashed lines), the prediction with the $\Lambda$CDM model in magenta. To complement the study, we plot the observations of SNIa from SCP2.1 in black points with their corresponding errors.\newline
The predictions for the distance modulus of $\Lambda$CDM and the EDE models lay quite close, especially at high redshift, and both are below the observations from $z \sim$ 1. Interestingly, both models fit very well at low redshift, when the luminosity distance grows linearly with redshift, independently of the model chosen.\newline
It is worth mentioning that the $\Lambda$CDM standard model was originally fitted to the data with WMAP-7 cosmological parameters, but with current cosmology, and particularly, the value for $H_0$, there is a slight discrepancy with SCP2.1 data at redshifts higher than 1.

\begin{figure}
	\includegraphics[width=\columnwidth]{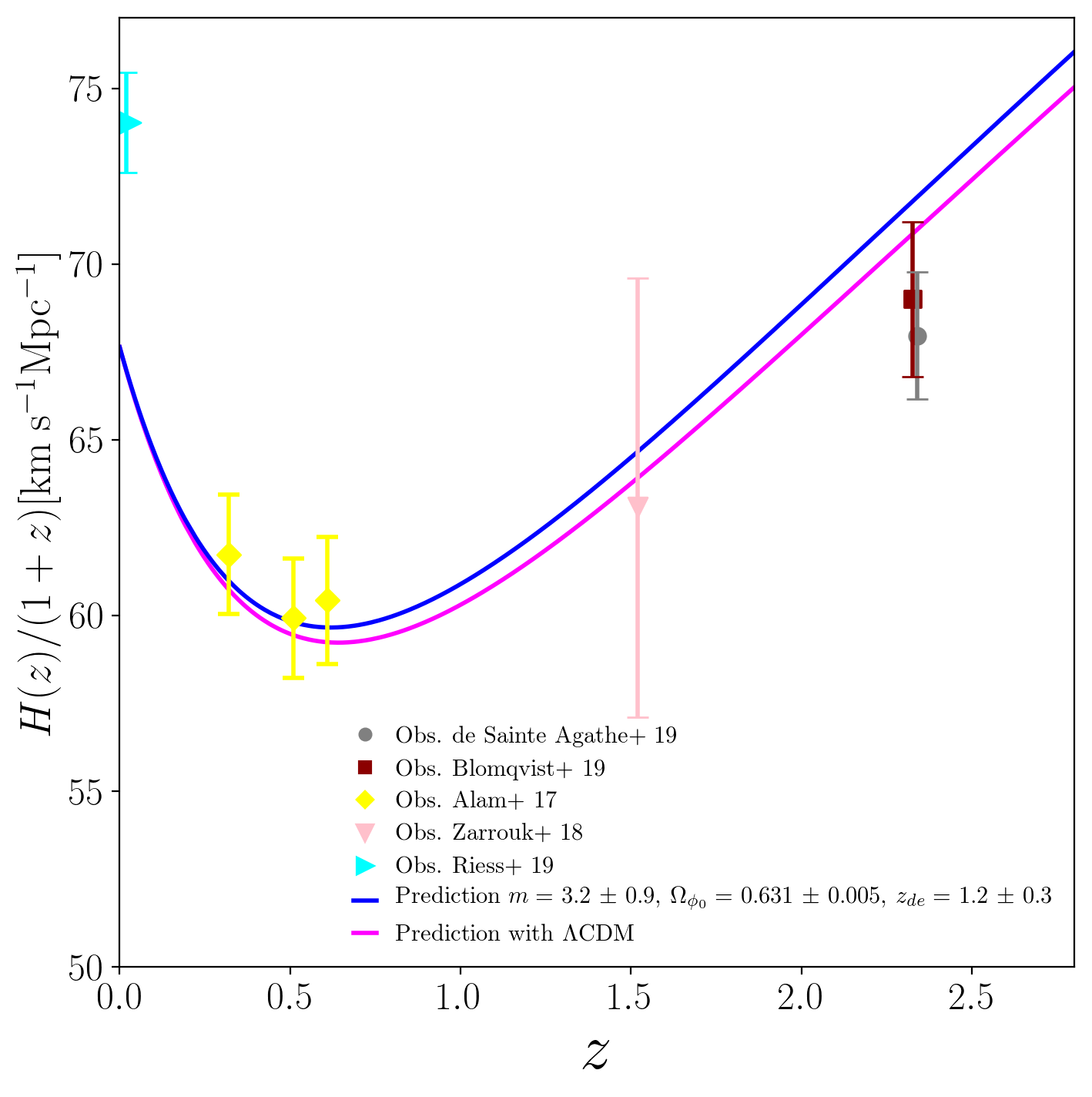}
\caption{Prediction for $H(z)/(1+z)$ as a function of $z$. We compare our model (blue solid line) with $\Lambda$CDM model (magenta solid line) and BAO observations derived with BOSS DR12 from \citet{alam2017} in yellow diamonds, from BOSS DR14 quasars by \citet{zarrouk2018} in the pink inverted triangle, BOSS DR14 Ly$\alpha$ autocorrelation at $z =$ 2.34 with the grey circle, and BOSS DR14 joint constraint from the Lya$\alpha$ auto-correlation and cross-correlation with quasars from \citet{blomqvist2019} in the dark red square. All the previous observations have computed with \citet{planckcol2018} cosmological parameters. The inferred Hubble measurement at $z =$ 0 derived independently by \citet{riess2019} is shown with the cyan right tilted triangle.}
\label{fig:hzvsz}
\end{figure}

Additional analysis is carried out with measurements of $H(z)/(1+z)$ vs. $z$ and our model prediction. Figure~\ref{fig:hzvsz} draws a comparison among our model (blue solid line) with BAO observations from BOSS DR12 from \citet{alam2017} in yellow diamonds, from BOSS DR14 quasars by \citet{zarrouk2018} in the pink inverted triangle, BOSS DR14 Ly$\alpha$ autocorrelation at $z =$ 2.34 with the grey circle, and BOSS DR14 joint constraint from the Lya$\alpha$ auto-correlation and cross-correlation with quasars from \citet{blomqvist2019} in the dark red square. All the previous observations have computed with \citet{planckcol2018} cosmological parameters. Finally, the inferred Hubble measurement today from \citet{riess2019} is shown with the cyan right tilted triangle. $\Lambda$CDM is plotted as a reference in a magenta line.\newline

\section{Standard Cosmological Probes}
\label{sec:5}
\subsection*{Age of the Universe with this model}

The age of the Universe for a given model in a standard cosmology is given by the expression:\newline
\begin{equation}\label{expression}
t_0 =\frac{1}{H_0}\int_0^{\infty} \frac{dz^{\prime}}{(1+z^{\prime})\sqrt{\Omega_{\phi 0}\cdot  f(z) + \Omega_{m 0} \cdot  (1+z)^{3}}},
\end{equation}
\noindent In our model, with the parameters in Table~\ref{tab:table}, the calculated age of the Universe is:\newline
\begin{equation}\label{age}
t_0 = 13.441 \pm 0.004 \:\:\: \text{Gyr}.
\end{equation}
\noindent As an important remark, \eqref{age} is only an approximation of the age of the Universe today, since the parametrization~\eqref{fun} needs further constraints with high redshift observables. However, the result is quite outstanding, taking into account that our model differs significantly at high redshift from the standard one.\newline

One way that the result can be interpreted is that the existence of early dark energy makes the Universe evolve faster than in the standard model. In this picture, the more negative $\omega$ is, the more accelerated is the expansion. Also, the Universe is ``younger'' if the dark energy component is precisely the one here proposed, given a value of $H_0$. 

\subsection*{\textit{Statefinder} parameters}

\begin{figure}
	\includegraphics[width=\columnwidth]{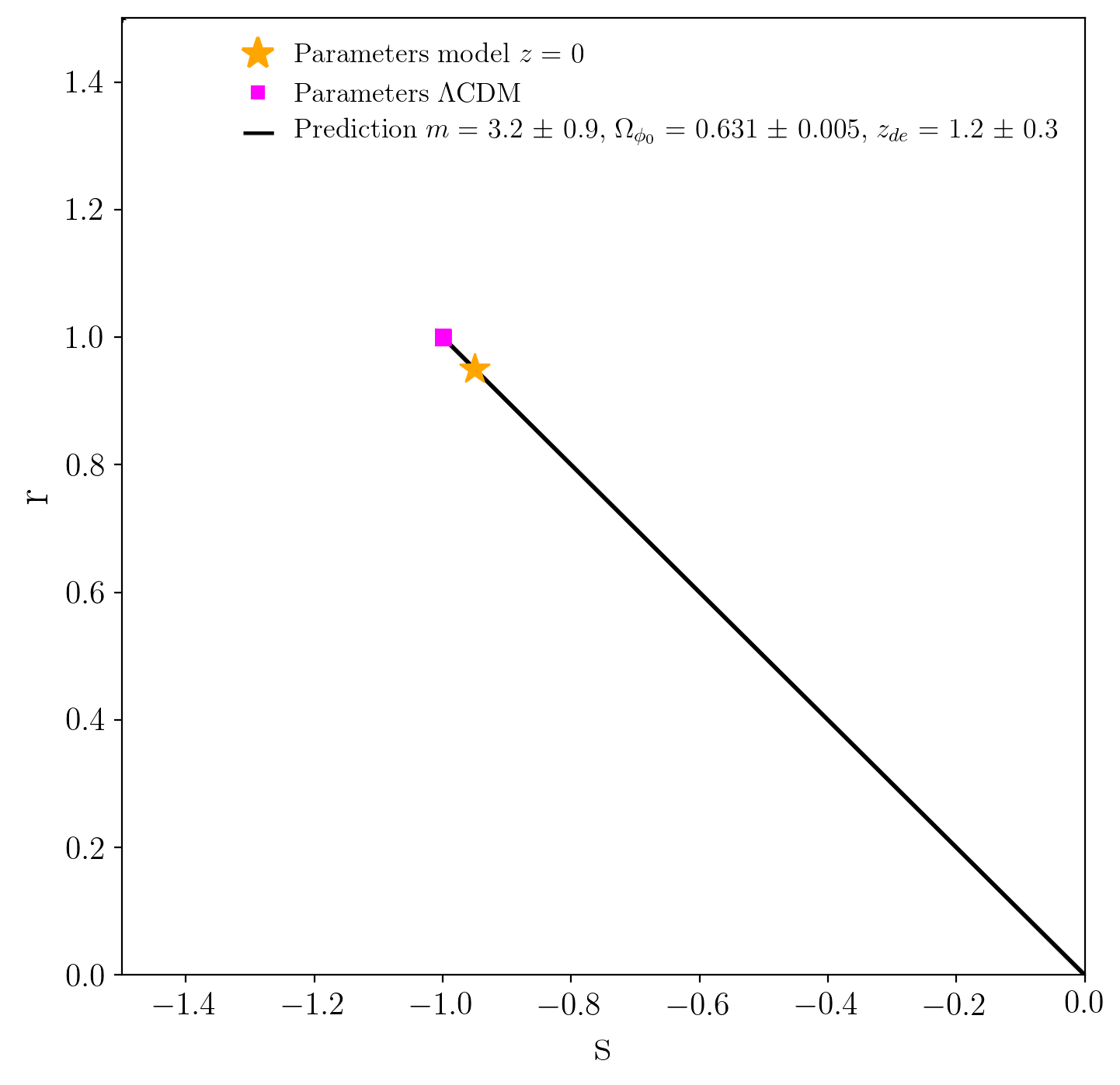}
\caption{\textit{Statefinder} parameters space. The set of parameters today for $\Lambda$CDM is presented with a purple square $\{r,s\} = \{1,-1\}$, while the values for our DE model are shown with the golden star. The black line exhibits the evolution with redshift of the \textit{Statefinder} parameters given our model. The effective parametrization evolves from high redshift (early times) in the right lower side to the future in the left upper corner.}
\label{fig:rs}
\end{figure}

In order to distinguish between a dark energy model and $\Lambda$CDM, \citet{gao2010} proposes a test using the \textit{Sta\-te\-fin\-der} parameters, defined as:
\begin{align}\label{rsprime}
r&=1+\frac{9}{2}\Omega_{\phi}\omega_{\phi}(1+\omega_{\phi})-\frac{3}{2}\Omega_{\phi}\frac{\dot{\omega_{\phi}}}{H}, \\
s&=1+\omega_{\phi}-\frac{1}{3}\frac{\dot{\omega_{\phi}}}{H\omega_{\phi}},
\end{align}
\noindent The values of these parameters with the Standard model are $\{r,s\} = \{1,-1\}$ today (i.e. at $z =$ 0). Any DE model-parameters will differ from the $\Lambda$CDM, and the departure of the former and the latter models in the space parameter at $z =$ 0 determines how extreme a DE model is compared with the behaviour of the cosmological cons\-tant.\newline
Figure~\ref{fig:rs} shows the \textit{Statefinder} space. The purple square and the golden star represent the $\Lambda$CDM and our DE model today, respectively. The black line shows the evolution of our model from the past ($\{r,s\} = \{0,0\}$) to the future ($r >$0 and $s <$0). Interestingly, the line evolves towards the prediction of the standard model at $z =$ 0, however, the equation of state has not reached the value $\omega_{\phi} = -$ 1 yet, therefore, there is still a gap between the refereed points in Figure~\ref{fig:rs}. In other words, our model is slightly off from the $\Lambda$CDM, because the prediction of the values $\Omega_{m 0}$, $\Omega_{\phi 0}$ and $z_*$ slightly differ from the standard model. Nevertheless, as discussed along this section, the equation of state relaxes and tends to $\Lambda$CDM model, once the field reaches the de-Sitter era.\newline

\section{Conclusions and perspectives}
\label{sec:6}

We have proposed a model of dark energy that causes an accelerated expansion of the Universe at late times, but also, has a non-negligible contribution during the radiation domination epoch. This dark energy candidate evolves from a radiation domination era to the De-Sitter time and emulates the behaviour of the cosmological constant. The properties of the dark energy component and it evolution in time (or redshift) have been extensively discussed, using an effective parametrization of the equation of state of the perfect fluid that could describe a scalar field.\newline

Using distance modulus of SNIa up to $z \sim$ 1.4 from the Supernova Cosmology Project 2.1 data sample, along with the CMB shift parameter $R_{\text{CMB}}$ and, the condition of the deceleration parameter equals to zero at $z = z_{de}$, we constrained the free parameters of our model: $\Omega_{\phi_0}$, the dark energy density of the field today, $m$, a factor that modules the transition between the radiation to the dark energy domination era, and, $z_{de}$, the redshift when the Universe reaches the De-Sitter era and its energy density overtakes the matter density (ending up the matter domination era). \newline
The complete solution of the parametrization allows us to study the dynamical evolution of the equation of state and the associated energy density fraction of the EDE candidate. Also, with the proposed method, we break the inner degeneracies among the free parameters.\newline

Ongoing work will impose additional constraints on the model by computing the energy density of the field during radiation, when the Universe is about a few minutes old, to study Big Bang Nucleosynthesis (BBN) and inferred parameters at this time. BBN is a well defined cosmological probe that can be used to rule out alternative models of dark matter and energy. Our model would not struggle in this cosmological regime since its energy contribution during radiation domination era is quite a subdominant, but non-negligible, therefore, it can play an important role as an effective degree of freedom of energy in the Hubble factor.\newline

Future efforts will be also focused on the most general family of solutions for the equation of state $\omega (z)$, using Heaviside step functions, that move between the cosmological domination epochs, and in the case of study, from the radiation to the De-Sitter era, satisfying different observational proxies. One of the crucial questions that arise with the introduction of these kind of equations for $\omega_{\phi}$ is the nature of dark energy and the interpretation of the energy density associated with the field $\rho_{de} (z) = \rho_{de0} \text{exp} [\int_{0}^{z} \frac{3}{1+z^{\prime}} (1+ \omega(z^{\prime})) dz^{\prime}]$. \newline

Finally, our goal is to fully understand if these alternative models for dark energy are competitive candidates to explain the accelerated expansion of the Universe at late times, avoiding the discrepancies that appear with the cosmological constant $\Lambda$, the fine-tuning after inflation and an unnecessary number of free parameters that have no physical interpretation. Our model has shown to provide compelling results as an early dark energy model. Ultimately, it seems about natural to have an evolving equation of state in the cosmological context, hence this study makes progress in this direction.

\section*{Acknowledgments}
This work was partly supported by the Observatorio Astron\'omico Nacional from Universidad Nacional de Colombia. L.A. Garc\'ia thanks Universidad ECCI for its funding.

\bibliographystyle{cas-model2-names}

\bibliography{mybibfile}

\begin{thebibliography}{41}
\expandafter\ifx\csname natexlab\endcsname\relax\def\natexlab#1{#1}\fi
\providecommand{\url}[1]{\texttt{#1}}
\providecommand{\href}[2]{#2}
\providecommand{\path}[1]{#1}
\providecommand{\DOIprefix}{doi:}
\providecommand{\ArXivprefix}{arXiv:}
\providecommand{\URLprefix}{URL: }
\providecommand{\Pubmedprefix}{pmid:}
\providecommand{\doi}[1]{\href{http://dx.doi.org/#1}{\path{#1}}}
\providecommand{\Pubmed}[1]{\href{pmid:#1}{\path{#1}}}
\providecommand{\bibinfo}[2]{#2}
\ifx\xfnm\relax \def\xfnm[#1]{\unskip,\space#1}\fi
\bibitem[{{Alam} et~al.(2017){Alam}, {Ata} and {Bailey}}]{alam2017}
\bibinfo{author}{{Alam}, S.}, \bibinfo{author}{{Ata}, M.},
  \bibinfo{author}{{Bailey}, Stephen, e.a.}, \bibinfo{year}{2017}.
\newblock \bibinfo{title}{{The clustering of galaxies in the completed SDSS-III
  Baryon Oscillation Spectroscopic Survey: cosmological analysis of the DR12
  galaxy sample}}.
\newblock \bibinfo{journal}{mnras} \bibinfo{volume}{470},
  \bibinfo{pages}{2617--2652}.
\newblock \DOIprefix\doi{10.1093/mnras/stx721},
  \href{http://arxiv.org/abs/1607.03155}{\tt arXiv:1607.03155}.
\bibitem[{{Alpher} et~al.(1948){Alpher}, {Bethe} and {Gamow}}]{alpher1948}
\bibinfo{author}{{Alpher}, R.A.}, \bibinfo{author}{{Bethe}, H.},
  \bibinfo{author}{{Gamow}, G.}, \bibinfo{year}{1948}.
\newblock \bibinfo{title}{{The Origin of Chemical Elements}}.
\newblock \bibinfo{journal}{Physical Review} \bibinfo{volume}{73},
  \bibinfo{pages}{803--804}.
\bibitem[{{Armend{\'a}riz-Pic{\'o}n} et~al.(1999){Armend{\'a}riz-Pic{\'o}n},
  {Damour} and {Mukhanov}}]{armendariz1999}
\bibinfo{author}{{Armend{\'a}riz-Pic{\'o}n}, C.}, \bibinfo{author}{{Damour},
  T.}, \bibinfo{author}{{Mukhanov}, V.}, \bibinfo{year}{1999}.
\newblock \bibinfo{title}{{k-Inflation}}.
\newblock \bibinfo{journal}{Physics Letters B} \bibinfo{volume}{458},
  \bibinfo{pages}{209--218}.
\newblock \href{http://arxiv.org/abs/hep-th/9904075}{\tt arXiv:hep-th/9904075}.
\bibitem[{{Armendariz-Picon} et~al.(2001){Armendariz-Picon}, {Mukhanov} and
  {Steinhardt}}]{armendariz2001}
\bibinfo{author}{{Armendariz-Picon}, C.}, \bibinfo{author}{{Mukhanov}, V.},
  \bibinfo{author}{{Steinhardt}, P.J.}, \bibinfo{year}{2001}.
\newblock \bibinfo{title}{{Essentials of k-essence}}.
\newblock \bibinfo{journal}{prd} \bibinfo{volume}{63}, \bibinfo{pages}{103510}.
\newblock \href{http://arxiv.org/abs/astro-ph/0006373}{\tt
  arXiv:astro-ph/0006373}.
\bibitem[{{Bento} et~al.(2004){Bento}, {Bertolami} and {Sen}}]{bento2004}
\bibinfo{author}{{Bento}, M.C.}, \bibinfo{author}{{Bertolami}, O.},
  \bibinfo{author}{{Sen}, A.A.}, \bibinfo{year}{2004}.
\newblock \bibinfo{title}{{Revival of the unified dark energy dark matter
  model?}}
\newblock \bibinfo{journal}{prd} \bibinfo{volume}{70}, \bibinfo{pages}{083519}.
\newblock \href{http://arxiv.org/abs/astro-ph/0407239}{\tt
  arXiv:astro-ph/0407239}.
\bibitem[{{Blomqvist} et~al.(2019){Blomqvist}, {du Mas des Bourboux} and
  {Busca}}]{blomqvist2019}
\bibinfo{author}{{Blomqvist}, M.}, \bibinfo{author}{{du Mas des Bourboux}, H.},
  \bibinfo{author}{{Busca}, Nicol{\'a}s~G., e.a.}, \bibinfo{year}{2019}.
\newblock \bibinfo{title}{{Baryon acoustic oscillations from the
  cross-correlation of Ly{\ensuremath{\alpha}} absorption and quasars in eBOSS
  DR14}}.
\newblock \bibinfo{journal}{aap} \bibinfo{volume}{629}, \bibinfo{pages}{A86}.
\newblock \DOIprefix\doi{10.1051/0004-6361/201935641},
  \href{http://arxiv.org/abs/1904.03430}{\tt arXiv:1904.03430}.
\bibitem[{{Bond} et~al.(1997){Bond}, {Efstathiou} and {Tegmark}}]{bond1997}
\bibinfo{author}{{Bond}, J.R.}, \bibinfo{author}{{Efstathiou}, G.},
  \bibinfo{author}{{Tegmark}, M.}, \bibinfo{year}{1997}.
\newblock \bibinfo{title}{{Forecasting cosmic parameter errors from microwave
  background anisotropy experiments}}.
\newblock \bibinfo{journal}{mnras} \bibinfo{volume}{291},
  \bibinfo{pages}{L33--L41}.
\newblock \href{http://arxiv.org/abs/astro-ph/9702100}{\tt
  arXiv:astro-ph/9702100}.
\bibitem[{{Caldwell}(2002)}]{caldwell2002}
\bibinfo{author}{{Caldwell}, R.R.}, \bibinfo{year}{2002}.
\newblock \bibinfo{title}{{A phantom menace? Cosmological consequences of a
  dark energy component with super-negative equation of state}}.
\newblock \bibinfo{journal}{Physics Letters B} \bibinfo{volume}{545},
  \bibinfo{pages}{23--29}.
\newblock \href{http://arxiv.org/abs/astro-ph/9908168}{\tt
  arXiv:astro-ph/9908168}.
\bibitem[{{Caldwell} et~al.(1998){Caldwell}, {Dave} and
  {Steinhardt}}]{caldwell1998}
\bibinfo{author}{{Caldwell}, R.R.}, \bibinfo{author}{{Dave}, R.},
  \bibinfo{author}{{Steinhardt}, P.J.}, \bibinfo{year}{1998}.
\newblock \bibinfo{title}{{Quintessential Cosmology Novel Models of
  Cosmological Structure Formation}}.
\newblock \bibinfo{journal}{apss} \bibinfo{volume}{261},
  \bibinfo{pages}{303--310}.
\bibitem[{{Carroll}(2001)}]{carroll2001}
\bibinfo{author}{{Carroll}, S.M.}, \bibinfo{year}{2001}.
\newblock \bibinfo{title}{{The Cosmological Constant}}.
\newblock \bibinfo{journal}{Living Reviews in Relativity} \bibinfo{volume}{4},
  \bibinfo{pages}{1}.
\newblock \href{http://arxiv.org/abs/astro-ph/0004075}{\tt
  arXiv:astro-ph/0004075}.
\bibitem[{{Chiba}(2002)}]{chiba2002}
\bibinfo{author}{{Chiba}, T.}, \bibinfo{year}{2002}.
\newblock \bibinfo{title}{{Tracking k-essence}}.
\newblock \bibinfo{journal}{prd} \bibinfo{volume}{66}, \bibinfo{pages}{063514}.
\newblock \href{http://arxiv.org/abs/astro-ph/0206298}{\tt
  arXiv:astro-ph/0206298}.
\bibitem[{{Chiba} et~al.(2000){Chiba}, {Okabe} and {Yamaguchi}}]{chiba2000}
\bibinfo{author}{{Chiba}, T.}, \bibinfo{author}{{Okabe}, T.},
  \bibinfo{author}{{Yamaguchi}, M.}, \bibinfo{year}{2000}.
\newblock \bibinfo{title}{{Kinetically driven quintessence}}.
\newblock \bibinfo{journal}{prd} \bibinfo{volume}{62}, \bibinfo{pages}{023511}.
\bibitem[{{Clifton} et~al.(2012){Clifton}, {Ferreira}, {Padilla} and
  {Skordis}}]{clifton2012}
\bibinfo{author}{{Clifton}, T.}, \bibinfo{author}{{Ferreira}, P.G.},
  \bibinfo{author}{{Padilla}, A.}, \bibinfo{author}{{Skordis}, C.},
  \bibinfo{year}{2012}.
\newblock \bibinfo{title}{{Modified gravity and cosmology}}.
\newblock \bibinfo{journal}{physrep} \bibinfo{volume}{513},
  \bibinfo{pages}{1--189}.
\newblock \DOIprefix\doi{10.1016/j.physrep.2012.01.001},
  \href{http://arxiv.org/abs/1106.2476}{\tt arXiv:1106.2476}.
\bibitem[{{Cline} et~al.(2004){Cline}, {Jeon} and {Moore}}]{cline2004}
\bibinfo{author}{{Cline}, J.M.}, \bibinfo{author}{{Jeon}, S.},
  \bibinfo{author}{{Moore}, G.D.}, \bibinfo{year}{2004}.
\newblock \bibinfo{title}{{The phantom menaced: Constraints on low-energy
  effective ghosts}}.
\newblock \bibinfo{journal}{prd} \bibinfo{volume}{70}, \bibinfo{pages}{043543}.
\newblock \href{http://arxiv.org/abs/hep-ph/0311312}{\tt arXiv:hep-ph/0311312}.
\bibitem[{{Davari} et~al.(2018){Davari}, {Malekjani} and
  {Artymowski}}]{davari2018}
\bibinfo{author}{{Davari}, Z.}, \bibinfo{author}{{Malekjani}, M.},
  \bibinfo{author}{{Artymowski}, M.}, \bibinfo{year}{2018}.
\newblock \bibinfo{title}{{New parametrization for unified dark matter and dark
  energy}}.
\newblock \bibinfo{journal}{prd} \bibinfo{volume}{97}, \bibinfo{pages}{123525}.
\bibitem[{{De Felice} and {Tsujikawa}(2010)}]{felice2010}
\bibinfo{author}{{De Felice}, A.}, \bibinfo{author}{{Tsujikawa}, S.},
  \bibinfo{year}{2010}.
\newblock \bibinfo{title}{{f( R) Theories}}.
\newblock \bibinfo{journal}{Living Reviews in Relativity} \bibinfo{volume}{13},
  \bibinfo{pages}{3}.
\newblock \href{http://arxiv.org/abs/1002.4928}{\tt arXiv:1002.4928}.
\bibitem[{{Doran} and {Robbers}(2006)}]{doran2006}
\bibinfo{author}{{Doran}, M.}, \bibinfo{author}{{Robbers}, G.},
  \bibinfo{year}{2006}.
\newblock \bibinfo{title}{{Early dark energy cosmologies}}.
\newblock \bibinfo{journal}{jcap} \bibinfo{volume}{2006}, \bibinfo{pages}{026}.
\bibitem[{{Efstathiou} and {Bond}(1999)}]{efstathiou1998}
\bibinfo{author}{{Efstathiou}, G.}, \bibinfo{author}{{Bond}, J.R.},
  \bibinfo{year}{1999}.
\newblock \bibinfo{title}{{Cosmic confusion: degeneracies among cosmological
  parameters derived from measurements of microwave background anisotropies}}.
\newblock \bibinfo{journal}{mnras} \bibinfo{volume}{304},
  \bibinfo{pages}{75--97}.
\newblock \href{http://arxiv.org/abs/astro-ph/9807103}{\tt
  arXiv:astro-ph/9807103}.
\bibitem[{Faraoni and Capozziello(2011)}]{capozziello2010}
\bibinfo{author}{Faraoni, V.}, \bibinfo{author}{Capozziello, S.},
  \bibinfo{year}{2011}.
\newblock \bibinfo{title}{{Beyond Einstein Gravity}}. volume
  \bibinfo{volume}{170}.
\newblock \bibinfo{publisher}{Springer}, \bibinfo{address}{Dordrecht}.
\newblock \URLprefix
  \url{http://www.springerlink.com/content/hl1805/#section=801705&page=1},
  \DOIprefix\doi{10.1007/978-94-007-0165-6}.
\bibitem[{{Gamow}(1946)}]{gamow1946}
\bibinfo{author}{{Gamow}, G.}, \bibinfo{year}{1946}.
\newblock \bibinfo{title}{{Expanding Universe and the Origin of Elements}}.
\newblock \bibinfo{journal}{Physical Review} \bibinfo{volume}{70},
  \bibinfo{pages}{572--573}.
\bibitem[{{Gao} and {Yang}(2010)}]{gao2010}
\bibinfo{author}{{Gao}, X.T.}, \bibinfo{author}{{Yang}, R.J.},
  \bibinfo{year}{2010}.
\newblock \bibinfo{title}{{Geometrical diagnostic for purely kinetic k-essence
  dark energy}}.
\newblock \bibinfo{journal}{Physics Letters B} \bibinfo{volume}{687},
  \bibinfo{pages}{99--102}.
\newblock \href{http://arxiv.org/abs/1003.2786}{\tt arXiv:1003.2786}.
\bibitem[{{Garc{\'{i}}a} et~al.(2011){Garc{\'{i}}a}, {Tejeiro} and
  {Casta{\~{n}}eda}}]{garcia2011}
\bibinfo{author}{{Garc{\'{i}}a}, L.}, \bibinfo{author}{{Tejeiro}, J.},
  \bibinfo{author}{{Casta{\~{n}}eda}, L.}, \bibinfo{year}{2011}.
\newblock \bibinfo{title}{{Primordial nucleosynthesis in the presence of
  sterile neutrinos}}.
\newblock \bibinfo{journal}{Proceedings of the International School of Physics
  Enrico Fermi} \bibinfo{volume}{178}, \bibinfo{pages}{309 -- 316}.
\bibitem[{{Gibbons}(2002)}]{gibbons2002}
\bibinfo{author}{{Gibbons}, G.W.}, \bibinfo{year}{2002}.
\newblock \bibinfo{title}{{Cosmological evolution of the rolling tachyon}}.
\newblock \bibinfo{journal}{Physics Letters B} \bibinfo{volume}{537},
  \bibinfo{pages}{1--4}.
\newblock \href{http://arxiv.org/abs/hep-th/0204008}{\tt arXiv:hep-th/0204008}.
\bibitem[{{Ho{\v r}ava} and {Minic}(2000)}]{horava2000}
\bibinfo{author}{{Ho{\v r}ava}, P.}, \bibinfo{author}{{Minic}, D.},
  \bibinfo{year}{2000}.
\newblock \bibinfo{title}{{Probable Values of the Cosmological Constant in a
  Holographic Theory}}.
\newblock \bibinfo{journal}{Physical Review Letters} \bibinfo{volume}{85},
  \bibinfo{pages}{1610--1613}.
\newblock \href{http://arxiv.org/abs/hep-th/0001145}{\tt arXiv:hep-th/0001145}.
\bibitem[{{Huang} et~al.(2015){Huang}, {Wang} and {Wang}}]{huang2015}
\bibinfo{author}{{Huang}, Q.G.}, \bibinfo{author}{{Wang}, K.},
  \bibinfo{author}{{Wang}, S.}, \bibinfo{year}{2015}.
\newblock \bibinfo{title}{{Distance priors from Planck 2015 data}}.
\newblock \bibinfo{journal}{jcap} \bibinfo{volume}{12}, \bibinfo{pages}{022}.
\newblock \href{http://arxiv.org/abs/1509.00969}{\tt arXiv:1509.00969}.
\bibitem[{{Kamenshchik} et~al.(2000){Kamenshchik}, {Moschella} and
  {Pasquier}}]{kamenshchik2000}
\bibinfo{author}{{Kamenshchik}, A.}, \bibinfo{author}{{Moschella}, U.},
  \bibinfo{author}{{Pasquier}, V.}, \bibinfo{year}{2000}.
\newblock \bibinfo{title}{{Chaplygin-like gas and branes in black hole bulks}}.
\newblock \bibinfo{journal}{Physics Letters B} \bibinfo{volume}{487},
  \bibinfo{pages}{7--13}.
\newblock \href{http://arxiv.org/abs/gr-qc/0005011}{\tt arXiv:gr-qc/0005011}.
\bibitem[{{Khoraminezhad} et~al.(2020){Khoraminezhad}, {Viel}, {Baccigalupi}
  and {Archidiacono}}]{ede2020}
\bibinfo{author}{{Khoraminezhad}, H.}, \bibinfo{author}{{Viel}, M.},
  \bibinfo{author}{{Baccigalupi}, C.}, \bibinfo{author}{{Archidiacono}, M.},
  \bibinfo{year}{2020}.
\newblock \bibinfo{title}{{Constraints on the Spacetime Dynamics of an Early
  Dark Energy Component}}.
\newblock \bibinfo{journal}{arXiv e-prints}
  \href{http://arxiv.org/abs/2001.10252}{\tt arXiv:2001.10252}.
\bibitem[{{Lorenz} et~al.(2017){Lorenz}, {Calabrese} and {Alonso}}]{lorenz2017}
\bibinfo{author}{{Lorenz}, C.S.}, \bibinfo{author}{{Calabrese}, E.},
  \bibinfo{author}{{Alonso}, D.}, \bibinfo{year}{2017}.
\newblock \bibinfo{title}{{Distinguishing between neutrinos and time-varying
  dark energy through cosmic time}}.
\newblock \bibinfo{journal}{prd} \bibinfo{volume}{96}, \bibinfo{pages}{043510}.
\newblock \href{http://arxiv.org/abs/1706.00730}{\tt arXiv:1706.00730}.
\bibitem[{{Panotopoulos}(2011)}]{panotopoulos2011}
\bibinfo{author}{{Panotopoulos}, G.}, \bibinfo{year}{2011}.
\newblock \bibinfo{title}{{A dynamical dark energy model with a given
  luminosity distance}}.
\newblock \bibinfo{journal}{General Relativity and Gravitation}
  \bibinfo{volume}{43}, \bibinfo{pages}{3191--3199}.
\newblock \href{http://arxiv.org/abs/1107.4475}{\tt arXiv:1107.4475}.
\bibitem[{{Peebles} and {Ratra}(2003)}]{peebles2003}
\bibinfo{author}{{Peebles}, P.J.}, \bibinfo{author}{{Ratra}, B.},
  \bibinfo{year}{2003}.
\newblock \bibinfo{title}{{The cosmological constant and dark energy}}.
\newblock \bibinfo{journal}{Reviews of Modern Physics} \bibinfo{volume}{75},
  \bibinfo{pages}{559--606}.
\newblock \href{http://arxiv.org/abs/astro-ph/0207347}{\tt
  arXiv:astro-ph/0207347}.
\bibitem[{{Planck Collaboration} et~al.(2015){Planck Collaboration}, {Ade},
  {Aghanim}, {Arnaud}, {Ashdown}, {Aumont}, {Baccigalupi}, {Banday},
  {Barreiro}, {Bartlett} and et~al.}]{planck2015}
\bibinfo{author}{{Planck Collaboration}}, \bibinfo{author}{{Ade}, P.A.R.},
  \bibinfo{author}{{Aghanim}, N.}, \bibinfo{author}{{Arnaud}, M.},
  \bibinfo{author}{{Ashdown}, M.}, \bibinfo{author}{{Aumont}, J.},
  \bibinfo{author}{{Baccigalupi}, C.}, \bibinfo{author}{{Banday}, A.J.},
  \bibinfo{author}{{Barreiro}, R.B.}, \bibinfo{author}{{Bartlett}, J.G.},
  \bibinfo{author}{et~al.}, \bibinfo{year}{2015}.
\newblock \bibinfo{title}{{Planck 2015 results. XIII. Cosmological
  parameters}}.
\newblock \bibinfo{journal}{ArXiv e-prints}
  \href{http://arxiv.org/abs/1502.01589}{\tt arXiv:1502.01589}.
\bibitem[{{Planck Collaboration} et~al.(2018){Planck Collaboration}, {Aghanim}
  and {Akrami}}]{planckcol2018}
\bibinfo{author}{{Planck Collaboration}}, \bibinfo{author}{{Aghanim}, N.},
  \bibinfo{author}{{Akrami}, Y., e.a.}, \bibinfo{year}{2018}.
\newblock \bibinfo{title}{{Planck 2018 results. VI. Cosmological parameters}}.
\newblock \bibinfo{journal}{arXiv e-prints} ,
  \bibinfo{pages}{arXiv:1807.06209}\href{http://arxiv.org/abs/1807.06209}{\tt
  arXiv:1807.06209}.
\bibitem[{{Ratra} and {Peebles}(1988)}]{ratra1988}
\bibinfo{author}{{Ratra}, B.}, \bibinfo{author}{{Peebles}, P.J.E.},
  \bibinfo{year}{1988}.
\newblock \bibinfo{title}{{Cosmological consequences of a rolling homogeneous
  scalar field}}.
\newblock \bibinfo{journal}{prd} \bibinfo{volume}{37},
  \bibinfo{pages}{3406--3427}.
\bibitem[{{Riess} et~al.(2019){Riess}, {Casertano}, {Yuan}, {Macri} and
  {Scolnic}}]{riess2019}
\bibinfo{author}{{Riess}, A.G.}, \bibinfo{author}{{Casertano}, S.},
  \bibinfo{author}{{Yuan}, W.}, \bibinfo{author}{{Macri}, L.M.},
  \bibinfo{author}{{Scolnic}, D.}, \bibinfo{year}{2019}.
\newblock \bibinfo{title}{{Large Magellanic Cloud Cepheid Standards Provide a
  1\% Foundation for the Determination of the Hubble Constant and Stronger
  Evidence for Physics beyond {\ensuremath{\Lambda}}CDM}}.
\newblock \bibinfo{journal}{apj} \bibinfo{volume}{876}, \bibinfo{pages}{85}.
\newblock \DOIprefix\doi{10.3847/1538-4357/ab1422},
  \href{http://arxiv.org/abs/1903.07603}{\tt arXiv:1903.07603}.
\bibitem[{{Riess} et~al.(2000){Riess}, {Filippenko} et~al.}]{riess2000}
\bibinfo{author}{{Riess}, A.G.}, \bibinfo{author}{{Filippenko}, A.V.}, et~al.,
  \bibinfo{year}{2000}.
\newblock \bibinfo{title}{{Tests of the Accelerating Universe with
  Near-Infrared Observations of a High-Redshift Type IA Supernova}}.
\newblock \bibinfo{journal}{apj} \bibinfo{volume}{536},
  \bibinfo{pages}{62--67}.
\newblock \href{http://arxiv.org/abs/astro-ph/0001384}{\tt
  arXiv:astro-ph/0001384}.
\bibitem[{{Rubin} et~al.(2014){Rubin}, {Aldering} et~al.}]{rubin2014}
\bibinfo{author}{{Rubin}, D.}, \bibinfo{author}{{Aldering}, G.S.}, et~al.,
  \bibinfo{year}{2014}.
\newblock \bibinfo{title}{{Updates to the Union SNe Ia Compilation}}, in:
  \bibinfo{booktitle}{American Astronomical Society Meeting Abstracts \#223},
  p. \bibinfo{pages}{245.09}.
\bibitem[{{Sami} and {Padmanabhan}(2003)}]{sami2003}
\bibinfo{author}{{Sami}, M.}, \bibinfo{author}{{Padmanabhan}, T.},
  \bibinfo{year}{2003}.
\newblock \bibinfo{title}{{Viable cosmology with a scalar field coupled to the
  trace of the stress tensor}}.
\newblock \bibinfo{journal}{prd} \bibinfo{volume}{67}, \bibinfo{pages}{083509}.
\newblock \href{http://arxiv.org/abs/hep-th/0212317}{\tt arXiv:hep-th/0212317}.
\bibitem[{{Sen}(2002a)}]{sen2002a}
\bibinfo{author}{{Sen}, A.}, \bibinfo{year}{2002}a.
\newblock \bibinfo{title}{{Rolling Tachyon}}.
\newblock \bibinfo{journal}{Journal of High Energy Physics}
  \bibinfo{volume}{4}, \bibinfo{pages}{048}.
\newblock \href{http://arxiv.org/abs/hep-th/0203211}{\tt arXiv:hep-th/0203211}.
\bibitem[{{Sen}(2002b)}]{sen2002b}
\bibinfo{author}{{Sen}, A.}, \bibinfo{year}{2002}b.
\newblock \bibinfo{title}{{Tachyon Matter}}.
\newblock \bibinfo{journal}{Journal of High Energy Physics}
  \bibinfo{volume}{7}, \bibinfo{pages}{065}.
\newblock \href{http://arxiv.org/abs/hep-th/0203265}{\tt arXiv:hep-th/0203265}.
\bibitem[{{Wetterich}(2004)}]{wetterich2004}
\bibinfo{author}{{Wetterich}, C.}, \bibinfo{year}{2004}.
\newblock \bibinfo{title}{{Phenomenological parameterization of quintessence}}.
\newblock \bibinfo{journal}{Physics Letters B} \bibinfo{volume}{594},
  \bibinfo{pages}{17--22}.
\bibitem[{{Zarrouk} et~al.(2018){Zarrouk}, {Burtin} and
  {Gil-Mar{\'\i}n}}]{zarrouk2018}
\bibinfo{author}{{Zarrouk}, P.}, \bibinfo{author}{{Burtin}, E.},
  \bibinfo{author}{{Gil-Mar{\'\i}n}, H{\'e}ctor, e.a.}, \bibinfo{year}{2018}.
\newblock \bibinfo{title}{{The clustering of the SDSS-IV extended Baryon
  Oscillation Spectroscopic Survey DR14 quasar sample: measurement of the
  growth rate of structure from the anisotropic correlation function between
  redshift 0.8 and 2.2}}.
\newblock \bibinfo{journal}{mnras} \bibinfo{volume}{477},
  \bibinfo{pages}{1639--1663}.
\newblock \DOIprefix\doi{10.1093/mnras/sty506},
  \href{http://arxiv.org/abs/1801.03062}{\tt arXiv:1801.03062}.

\end{thebibliography}


\end{document}